\documentclass[a4paper]{article}

\usepackage{INTERSPEECH2019}
\usepackage{amsmath,graphicx,subfigure,hyperref}
\usepackage{tabularx, xcolor}

\title{Dereverberation using joint estimation of dry speech signal and acoustic system}
\name{Sanna Wager$^{1*, 2}$, Keunwoo Choi$^1$, Simon Durand$^1$}

\address{
  $^1$Spotify, S.A., New York\\
  $^2$Indiana University Bloomington, Luddy School of Informatics, Computing, and Engineering\thanks{*The research work described in this report was supported by the internship program at Spotify, S.A., in collaboration with the music intelligence team.}}
\email{scwager@indiana.edu, keunwooc@spotify.com, durand@spotify.com}

\begin{document}

\maketitle
\begin{abstract}
The purpose of speech dereverberation is to remove quality-degrading effects of a time-invariant impulse response filter from the signal. In this report, we describe an approach to speech dereverberation that involves joint estimation of the dry speech signal and of the room impulse response. We explore deep learning models that apply to each task separately, and how these can be combined in a joint model with shared parameters.
\end{abstract}


\section{Introduction}
\label{sec:intro}
Room acoustics can cause degradation of audio quality in recordings produced outside of professional recording studios. Acoustics in everyday environments such as the home commonly produce undesirable effects including excessive coloration from early reflections and masking from late reflections \cite{faller2019modifying}. One may wish to mitigate these effects by dereverberating the speech signals. This type of speech enhancement can be applied in the post-processing stage, where the full signal has been recorded in advance, or in real time, if low latency is required. This technical report addresses the former case.

We focus on the single-channel case, where only one microphone was used to record the speech signal. The single-channel case is challenging because, unlike in the multi-channel case, a reverberation reduction program cannot utilize the difference in time of arrivals of reflections across the microphones. In the recent REVERB challenge \cite{kinoshita2016summary}, only one among eleven submitted single-channel systems both reduced the perceived amount of reverberation and improved overall quality.

The task of reverberation reduction has been framed in multiple ways over the past five decades. Habets \cite{habets2016fifty} describes three trends. The first involves modeling the acoustic system and using this information to equalize or filter the reverberant signal. The second approach is similar to denoising, treating the source and reverberation signals as independent. The third directly estimates the dry signal without trying to model the room acoustics. 

The techniques in the first trend, which involves modeling the acoustic system, provide clean results under specific conditions \cite{neely1979invertibility} and can provide the ability to control the Room Impulse Response (RIR) \cite{habets2016fifty}. However, they can be challenging to develop even when the RIR is known. Though the RIR function is a linear transformation, in realistic room acoustics, an exact inverse is often either unstable or acausal. Exact equalization requires very long filters, causing increased computational complexity and numerical instability. Sometimes, computing the inverse can be intractable \cite{neely1979invertibility, cecchi2018room}. Another issue is that imperfections in the RIR measurement will affect the result. Faller describes how inversion-based filtering only works in the ``sweet spot'' \cite{faller2019modifying}. Movement from the ``sweet spot'' to other positions in the room degrades the quality. Furthermore, changes in object positions in the room, temperature, or humidity will change the RIR, which can be described as a weakly non-stationary process \cite{cecchi2018room}. The RIR estimate then needs to be updated across the signal instead of only being computed once. For these reasons, techniques that remove only the most problematic components of the reverberation while leaving the rest intact have often been chosen for practical uses \cite{cecchi2018room}.

Approaches involving denoising or source separation do not necessarily generalize to the dereverberation task. The generalization is most problematic in the case of models where the target and noise signals are assumed independent---for example, time-frequency bin masking \cite{hershey2016deep}. Masking works best when bins belong either to the signal or to the noise, and there is not much overlap. A reverberant signal, though, is a sum of delayed and filtered versions of the original signal, which are highly correlated and likely present in the same time-frequency bins. Another assumption commonly made for denoising and that does not apply to dereverberation is that every time frame can be treated separately. This assumption also does not generalize well because RIRs often last longer than one Short-Time Fourier Transform (STFT) window.

Direct estimation approaches, e.g., \cite{ernst2018speech}, focus instead on estimating a clean signal given a reverberant input. They do not directly estimate room acoustics, or separate the dry and reverberant signals. Direct estimation can produce good results when the model is able to represent the distribution of the target outputs---speech signals. However, the nature of the direct estimation approach means that some useful information about the room acoustics is not utilized, which could negatively affect the model's ability, for example, to adapt to unseen room conditions.

In this report, we conduct preparatory work for developing a joint model, which combines the direct estimation approach with estimation of the RIR. A joint model, which involves shared parameters for these two tasks, can produce higher accuracy than separate models. We utilize the fact that we have access both to pairs of dry and reverberant signals and the RIRs used to generate these pairs via convolution. Access to the target dry speech and RIR signals allows us to train both components of the model in a supervised manner. In this report, we explore separate models for each task, and provide an overview of how these models can be combined into a single one. 

Section \ref{sec:related} covers related work. In section \ref{sec:models}, we provide an overview of the joint training technique. We also describe the models we use separately for the tasks of dry speech estimation and RIR estimation. In section \ref{sec:dataset}, we describe our experiments along with the dataset and pre-processing steps. Finally, section \ref{sec:conclusion} concludes the report.

\section{Related work}
\label{sec:related}
Deep neural networks (DNNs) have been investigated for the reverberation reduction task. Fully connected models that learn and process the signal frame-by-frame are proposed in \cite{han2015learning, williamson2017speech}. The first model outputs a magnitude STFT, while the second outputs both real and complex values. Williamson \textit{et al.} \cite{williamson2017speech} predict one frame at a time from an input of multiple consecutive frames. Each input frame includes a set of different time-frequency features extracted from the audio.

Ori \textit{et al.} \cite{ernst2018speech} investigate a fully convolutional model---a U-net \cite{ronneberger2015u}---to estimate dry speech directly from reverberant and noisy speech. The authors predict the magnitude spectrogram and use the reverberant phase for reconstruction of the audio signal. Results are state-of-the-art. In addition to supervised training, the authors experiment with using Generative Adversarial Networks (GANs) to improve over their supervised U-net model, but their performance decreases slightly.
    
GANs for the dereverberation task have also been proposed by Li \textit{et al.} \cite{li2018single}. The authors experiment with supervised models and further train the one that performed the best---a Long-Short-Term Memory Unit (LSTM) \cite{hochreiter1997long} using adversarial training. The downstream task in the work, however, is automatic character recognition instead of audio signal enhancement.
    
Engel \textit{et al.} \cite{engel2020ddsp} introduce a DSP-based DNN that includes differentiable oscillators, filters, and reverberation components, allowing for separate control over parameters such as pitch and loudness. The model has an encoder-decoder structure, and indicates potential for encoding the dry component of the signal while discarding the room reverberation, and reconstructing a dry signal, optionally convolving it with a different RIR.
    
We also mention relevant models designed for related tasks such as speech denoising. Tan \textit{et al.} \cite{tan2018convolutional} introduce the convolutional recurrent neural network (CRNN) model. The CRNN combines the convolutional filter feature extraction capability with the sequential modeling of an LSTM in an encoder-decoder setup. Huang \textit{et al.} address the speech separation task, using joint training \cite{huang2015joint} to improve accuracy by estimating both the mask and the target speech signal. Relevant work also includes non-negative matrix factorization approaches \cite{mohammadiha2015joint}, which separately dereverberate magnitude STFTs over every frequency bin. This technique greatly reduces the number of parameters and inspires our approach for RIR prediction.

While prior work exists that involves estimation of RIR parameters such as the direct-to-reverberant ratio or T60 \cite{eaton2016estimation, bryan2019data}, we are not aware of work that directly predicts the RIR from the reverberant speech.

\section{Models}
\label{sec:models}
\subsection{Joint training}
\begin{figure}[t]
    \centering
    \includegraphics[width=\columnwidth]{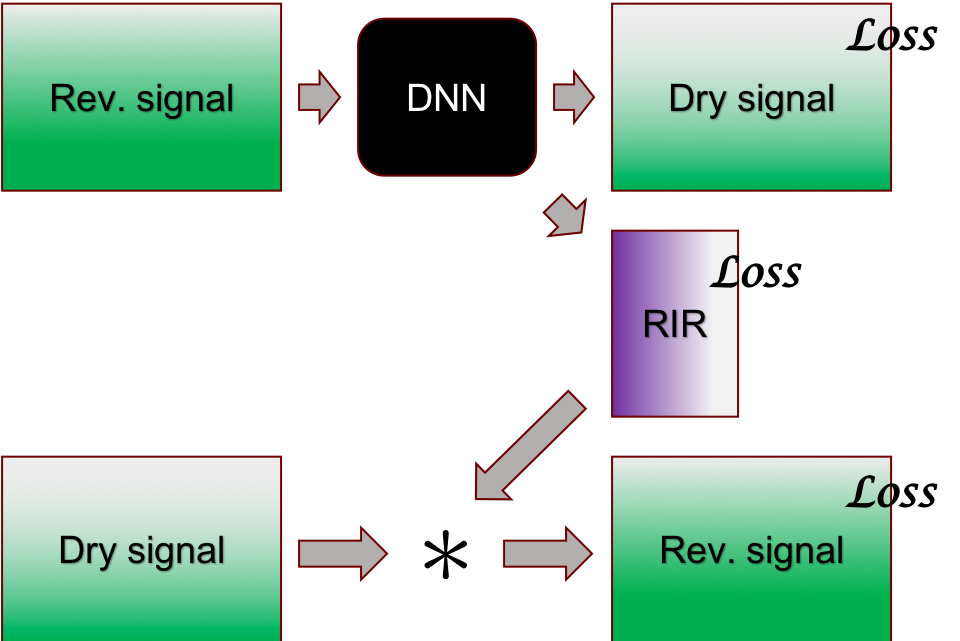}
    \caption{Overview of training technique. The DNN is trained to jointly estimate the dry signal and the RIR. Parameters are shared across the two tasks. The estimated RIR is convolved with the known target dry signal to produce a reverberant sigal estimate.}
    \label{fig:joint}
\end{figure}

Figure \ref{fig:joint} illustrates our joint training approach, which involves estimating both the dry signal and the RIR. Our assumption is that sharing some of the parameters used for the two tasks can improve accuracy compared to what the two models would reach when trained separately. The model would learn to extract the structure of the RIR from the reverberant signal, which would also make it better able to predict the dry signal. In addition to the dry speech and RIR training targets, we introduce a third target: The RIR estimate output by the model is convolved with the dry signal to produce a reconstruction of the reverberant signal. Given that we synthesize reverberant speech by convolving the dry signal and the RIR, all three targets are known and we can train the joint model in a supervised manner with three weighted loss components. In the following sections, we describe models we use separately for the two tasks. Once the separate models produce sufficiently accurate results, they can be combined into a joint model. The individual loss terms are then summed in a weighted manner.

\subsection{Direct estimation}
We first compare two existing models for estimating the dry speech directly from the reverberant signal. We modify the LSTM proposed in \cite{wang2018investigating} by using a Bi-directional Gated Recurrent Unit (Bi-GRU) \cite{chung2014empirical} instead. The model includes residual connections between layers. Given that we use a bi-directional model, we use half the hidden dimension used by Wang \textit{et al.}, which amounts to 380, but produces the original 760 dimensions when combining the two directions. 
We do not use exponential decay like the authors and solely rely on the learning rate decay adjusted by Adam optimizer \cite{kingma2014adam}. The target is also different as we estimate the log-STFT instead of recognizing characters. We empirically find that training parameters used by Wang \textit{et al.} produce the best results for our model.

We also implement the U-net introduced in \cite{ernst2018speech} for comparison.

\subsection{Room impulse response estimation}
We then explore using a fully convolutional model to estimate the RIR from the reverberant signal. We again choose to operate in the magnitude-frequency domain. We introduce a convolutional model designed to capture the time-invariant filter structure across a few seconds of the speech signal. It includes filters that span multiple time frames but only one frequency bin at a time, assuming the reverberation can be isolated per-frequency because the reverberant signal is a linear sum of the dry signal. The structure is as follows, with the first two values indicating the filter size (time and frequency axes) and the third the number of filters: (9$\times$1, 16), (14$\times$1, 32), (27$\times$1, 64), (27$\times$1, 32), (27$\times$1, 16), (28$\times$1, 4), (187$\times$1, 126). Each convolution is followed by an exponential linear unit activation \cite{clevert2015fast} except for the final one, which is followed by a rectified linear unit activation \cite{nair2010rectified}. After the final layer, the filter axis with 126 dimension becomes the time axis of the RIR. We apply this modification as the time-varying structures of the speech and RIR signals are unrelated.

\section{Experimental results}
\subsection{Dataset}
\label{sec:dataset}
For RIRs, we use publicly available recordings downloaded using Kaldi scripts\footnote{\url{https://github.com/kaldi-asr/kaldi/blob/master/egs/aspire/s5/local/multi_condition/rirs/}}. Data sources consist of the Aachen Impulse Response Database \cite{jeub2009binaural_aachen_air}, the C4DM Room Impulse Response Dataset \cite{stewart2010database_c4dm}, the RWCP Sound Scene Database in Real Acoustical Environments \cite{nakamura2000acoustical_rwcp}, the REVERB Challenge dataset's RIRs (omitting the noise signals) \cite{kinoshita2013reverb}, and the PORI Concert Hall Impulse Responses \cite{merimaa2005concert}. These combined provide a total of 1069 RIRs. Additionally, Steinmetz kindly provided RIRs used for research on a NeuralReverberator\footnote{\url{https://www.christiansteinmetz.com/projects-blog/neuralreverberator}}. This set includes 693 RIRs from a variety of sources.

We split the RIRs into 1362, 200, and 200 samples for training, validation, and testing, respectively. We group together signals that were too similar, e,g., involved the same source in the same room, the same receiver in same room, or the same RIR recorded with different microphones or microphone rotations. When the configurations were unknown, we group RIRs by the codes found in the file names. We assign groups larger than 20 to the training set for more balanced validation and test sets. We also limit the size of each group to 100 and discard the remaining RIRs. This data preparation results in a total 823 groups distributed across training, validation, and test sets.

\subsection{Format and pre-processing}
We process the input signals using a sample rate of 16000 Hz. We convolve each dry speech signal with multiple, randomly selected RIRs. We align the dry and reverberant speech in time by applying the RIR delay to the dry speech. We remove the near-silent part of the signal at the beginning. Finally, we truncate or pad the speech signals so that they last 5 seconds. We also zero-pad the RIRs to a minimum length of 2 seconds but do not truncate longer ones. 

We compute the STFT with frame size 32 ms and hop length 16 ms. We normalize the audio by dividing each STFT by its maximum. 

\subsection{Current results}
\label{sec:experiments}
The U-net introduced by Ernst \textit{et al.} produced the best results for the dry speech direct estimation task. The Bi-GRU produced too many artifacts to be used in practice. We also explored fully connected models, but found that they were not as accurate as the U-net. For the RIR estimation task, the convolutional model output impulses that had the desired decay but were not precise enough. The convolutional model might not be powerful enough to capture the time-invariant filter from the speech signal. RIRs have an unusual structure, which might be difficult for a DNN to predict: a short and loud impulse followed by a rapid decay. Structures of signals often estimated using DNNs, including speech signals or time series, have a distribution that remains relatively stable over time. To properly estimate a RIR, the DNN needs to concentrate most of the output energy into a small area. 

We believe a time-domain approach can be used to improve the precision of the U-net by including phase into the estimation. We also expect to improve RIR estimation results by using a sequential model, such as an LSTM in the time-frequency domain or a CRNN in the time domain. The time-alignment of the dry and reverberant speech, along with the removal of any silence at the beginning of the RIR signal, can also be further improved for better precision.

\section{Conclusion}
\label{sec:conclusion}
In this technical report, we describe a joint modeling approach to estimating dry speech and the room impulse response from a reverberant speech input. We provide an overview of separate deep learning models for each task and how these can be combined in future work. We operate on magnitude-STFT data. We find that a magnitude-STFT U-net \cite{ernst2018speech} provides good results for direct speech estimation. For room impulse response estimation, we introduce a convolutional model and plan to improve it further by modeling the sequential nature of the data. Though there has been some prior work using representations that include phase, namely, time-domain signals \cite{luo2018real} and complex STFTs \cite{williamson2017speech}, we leave representations that include the phase for future work. 


\bibliographystyle{IEEEtran}

\bibliography{mybib}

\end{document}